\documentclass[conference]{IEEEtran}
\IEEEoverridecommandlockouts
\usepackage{cite}
\usepackage{amsmath,amssymb,amsfonts}
\usepackage{algorithmic}
\usepackage{graphicx}
\usepackage{textcomp}
\usepackage{xcolor}
\usepackage{xr}
\usepackage{booktabs}
\usepackage[skip=0.5em]{caption}
\usepackage{multirow}
\usepackage{orcidlink}
\usepackage{paralist}
\usepackage{adjustbox}
\usepackage{multirow}
\def\BibTeX{{\rm B\kern-.05em{\sc i\kern-.025em b}\kern-.08em
    T\kern-.1667em\lower.7ex\hbox{E}\kern-.125emX}}
\begin{document}

\title{LLMs-Powered Real-Time Fault Injection: An Approach Toward Intelligent Fault Test Cases Generation \\

}

\author{\IEEEauthorblockN{1\textsuperscript{st} Mohammad Abboush \orcidlink{0000-0002-5533-0029}}
\IEEEauthorblockA{\textit{Technische Universität Clausthal} \\
Clausthal-Zellerfeld, Germany \\
mohammad.abboush@tu-clausthal.de}
\and
\IEEEauthorblockN{2\textsuperscript{nd} Ahmad Hatahet \orcidlink{0009-0009-7677-7514}} 
\IEEEauthorblockA{\textit{Technische Universität Clausthal} \\
Clausthal-Zellerfeld, Germany \\
ahmad.hatahet@tu-clausthal.de}

\and
\IEEEauthorblockN{3\textsuperscript{th} Andreas Rausch \orcidlink{0000-0002-6850-6409}}
\IEEEauthorblockA{\textit{Technische Universität Clausthal} \\
Clausthal-Zellerfeld, Germany \\
andreas.rausch@tu-clausthal.de}

}

\maketitle

\begin{abstract}

A well-known testing method for the safety evaluation and real-time validation of automotive software systems (ASSs) is Fault Injection (FI). In accordance with the ISO 26262 standard, the faults are introduced artificially for the purpose of analyzing the safety properties and verifying the safety mechanisms during the development phase. However, the current FI method and tools have a significant limitation in that they require manual identification of FI attributes, including fault type, location and time. The more complex the system, the more expensive, time-consuming and labour-intensive the process. To address the aforementioned challenge, a novel Large Language Models (LLMs)-assisted fault test cases (TCs) generation approach for utilization during real-time FI tests is proposed in this paper.  
To this end, considering the representativeness and coverage criteria, the applicability of various LLMs to create fault TCs from the functional safety requirements (FSRs) has been investigated. Through the validation results of LLMs, the superiority of the proposed approach utilizing gpt-4o in comparison to other state-of-the-art models has been demonstrated. Specifically, the proposed approach exhibits high performance in terms of FSRs classification and fault TCs generation with F1-score of 88\% and  97.5\%, respectively. To illustrate the proposed approach, the generated fault TCs were executed in real time on a hardware-in-the-loop system, where a high-fidelity automotive system model served as a case study. This novel approach offers a means of optimizing the real-time testing process, thereby reducing costs while simultaneously enhancing the safety properties of complex safety-critical ASSs.
\end{abstract}

\begin{IEEEkeywords}
automotive software systems development (ASSs), real-time validation, fault injection, LLMs, safety assessment, hardware-in-the-loop (HIL).  
\end{IEEEkeywords}


\section{Introduction}


The rapid evolution of software-controlled features in automotive software systems (ASSs) has resulted in a significant expansion of the system architecture and an attendant increase in complexity \cite{abelein2012complexity}. Meanwhile, considering the increasing complexity of modern vehicles, with several hundred control units connected to heterogeneous components, the failure rate is also increasing \cite{garousi2018testing3}. To mitigate the effects of failures and ensure that safety objectives are met without compromise, the ISO 26262 functional safety standard has been introduced as a unified development approach \cite{pintard2014safety}. 

At the system integration and testing phase (in accordance with ISO 26262-Part 4) \cite{url}, fault injection (FI) test has been recommended to evaluate system reliability and safety in the context of abnormal operating conditions \cite{hsueh1997fault}. To perform the FI test, three different fault-related attributes should be defined, i.e., fault type, fault location and FI time. During the safety engineering process, the fault attributes are defined on the basis of the principle of establishing edge-case scenarios that could potentially occur in real-world environment. Nevertheless, the FI is conducted either randomly \cite{ubar2010parallel} or exhaustively \cite{zheng2009monte}, depending on the domain expertise of the practitioner. This, in turn, lead to either a reduction in the test coverage due to the limited number of representative fault test cases (TCs) \cite{natella2012fault}, or the creation of an unlimited number of potential fault TCs \cite{benso2003fault}. The higher the complexity of the target system, the greater the expense of implementation in terms of time, effort and difficulty. Therefore, It is necessary to develop an intelligent approach that is capable of generating realistic fault TCs in a systematic manner, based on the FSRs for real-time testing scenarios. 
The recent advancements in neural network architectures, computational resources, and availability of larger-scale datasets have paved the way for the evolution of large language models (LLMs) as a field within natural language processing and artificial intelligence (AI) \cite{zhao2023survey}. The capabilities of LLMs extend beyond the domains of question-answering, machine translation and text generation, encompassing also the capacity to follow supporting the software and systems engineers during the testing process \cite{wang2024softwar}. Among these models, GPT4o \cite{gpt4o}, LLaMA \cite{ollama}, phi4 \cite{phi4} and qwen2.5 \cite{qwen2} stand out in addressing various engineering challenges due to their billions of parameters and extensive pre-training on diverse textual datasets covering \cite{hou2024large}. 

This paper investigates the applicability of integrating LLMs into the FI approach for the generation of fault TCs from FSRs considering extendable list of sensors with both single and concurrent faults. Furthermore, in this study, the effect of the generated LLMs-based fault TCs on the system behavior during real-time validation on a hardware-in-the-loop (HIL) system is analyzed. To verify the effectiveness of the proposed framework, a high-fidelity gasoline engine and a vehicle dynamics models were considered as automotive use cases. To the best of our knowledge, this is the first study to investigate the integration of LLMs with the FI test during the real-time validation of ASSs on a HIL simulation system. 

The main contributions of the proposed work are summarized as the following: 
\begin{itemize}
    \item Proposing a novel approach for generating the fault TCs from the FSRs based on LLMs so that the real-time FI process can be optimized and performed efficiently.
    \item To efficiently generate the fault TCs from diverse FSRs, LLM-based classification task is performed to group the sensor-related FSRs and the actuator-related FSRs.
    \item On the basis of the classified textual FSRs, a representative fault TCs for real-time FI process on HIL system is accurately identified considering extendable list of sensors with both single and concurrent faults.
    \item A comprehensive analysis of various types of LLMs is performed considering the effectiveness of data size and type of LLMs on the performance of fault TCs generation.
    \item The real-time behavior of high-fidelity automotive system under injecting the generated fault TCs have been demonstrated and analyzed.
\end{itemize}


\section{Related work}

Several methodologies have been proposed to generate fault TCs from textual requirements during FI experiments. Vedder et al. \cite{vedder2013combining} combined property-based and FI testing, using QuickCheck and FaultCheck to derive fault TCs from system specifications, validated on an airbag system. Similarly, Cong et al. \cite{cong2015automatic} introduced a runtime FI-based approach for driver robustness testing, using a bounded trace-based iterative strategy to identify TCs that trigger functional violations.

With the emergence of machine learning (ML), new strategies have optimized FI testing by learning from data. Khosrowjerdi et al. \cite{khosrowjerdi2018virtualized} proposed an ML-assisted virtual FI framework to generate fault TCs from formal requirements, validated on ECU systems using a toolchain built on QEMU, GDB, and LBTest. However, these methods focus primarily on code-level systems. In contrast, our work integrates both the high-fidelity system model and its corresponding code into a real-time hardware-in-the-loop (HIL) setup.

At the system level, Sedaghatbaf et al. \cite{sedaghatbaf2022delfase} developed DELFASE, a deep learning framework using GANs to explore fault spaces and identify critical faults. While improving coverage over random sampling, it selects faults from a predefined space rather than generating them.

The rise of transformer-based LLMs has further advanced software testing. Garg et al. \cite{garg2024coupling} showed that µBERT can enhance mutant generation as TCs, achieving broad abnormal condition coverage. Khanfir et al. \cite{khanfir2023efficient} used CodeBERT to seed representative code-level faults, improving detection by 17\%. Cotroneo et al. \cite{cotroneo2024neural} integrated LLMs with reinforcement learning from human feedback (RLHF) to automate fault scenario creation and reduce manual effort.

While these approaches show promising results in optimizing FI, they remain limited to function-level mutant generation and lack real-time, system-level applicability in automotive contexts. In contrast, our study targets realistic sensor- and actuator-related faults, aligned with ISO~26262 standards, and leverages various LLMs to improve the efficiency, impact, and coverage of fault TCs in real-time safety validation of automotive systems.

\section{Proposed Framework}

\begin{figure*}[thb]

    \centering
	\includegraphics[trim={0.08cm 0.08cm 0.08cm 0.08cm}, clip,width=0.9\linewidth]{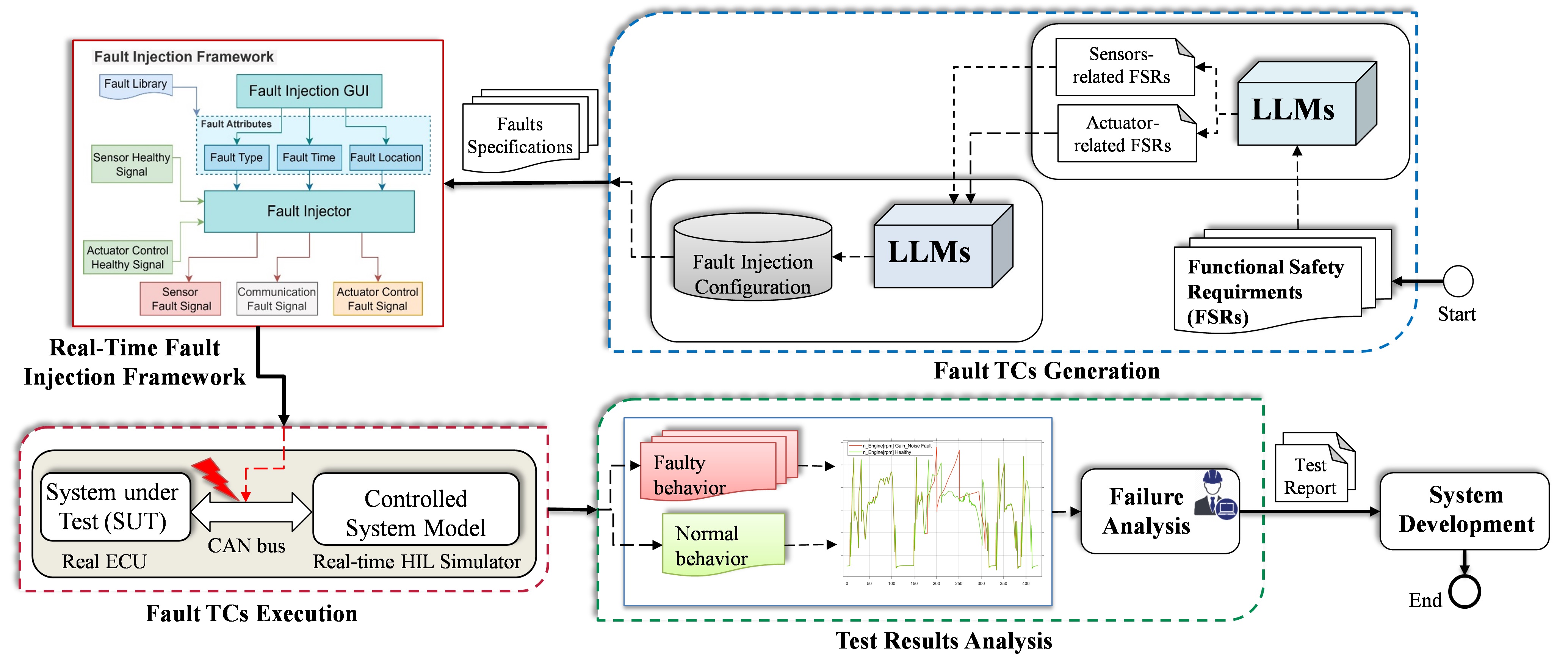}
	\caption{Proposed LLMs-assisted real-time fault injection approach with HIL simulation.}
	\label{fig:flowchart}       
\end{figure*}

The proposed LLMs-assisted real-time FI approach is shown in Figure~\ref{fig:flowchart}. The proposed approach consists of three phases, i.e., fault TCs generation, execution and result analysis, which constitute the process of real-time validation of ASSs using FI method.

In the initial phase, the automatic generation of fault TCs is initiated by feeding the LLMs with FSRs as text input. It is noteworthy that, in contrast to traditional ML models, the preprocessing of textual inputs is not required in this phase. This is due to the fact that LLMs utilise its inherent processing capabilities to process textual information as a natural language. The FSRs represents the criteria and guidelines by which the system is ensured to perform its intended functional safety under both normal and fault conditions. Some examples of the FSRs can be found in the following statement: 
\begin{itemize}
    \item FSR1: "In case of uncertainty in the vehicle velocity data, the SUT shall automatically adjust the speed of the vehicle to maintain a safe following distance of at least five meters from the vehicle ahead"
    \item FSR2: "In the event of simultaneous failures in both the throttle and brake actuators, the system shall activate an emergency deceleration procedure and automatically engage hazard lights".
\end{itemize}

Subsequently, the required specifications are evaluated and the most important system characteristics are determined by LLMs. This process ultimately results in the definition of the target functionality, e.g., emergency braking.

It is important to note that the FSRs are diverse and randomly distributed, concentrating on various system components, e.g., sensors, actuators and ECU. To address this complexity, a classification task is performed by the LLMs to categorise the FSRs based on the target components, e.g., sensor-related FSRs and actuator-related FSRs. By focusing on the specific class, the generation performance of fault TCs can be improved, thus reducing uncertainty.

Once the FSRs are classified based on the target components, the potential faults that could result in a violation of the determined functionality are identified. The fault categories may encompass a multitude of aspects within the system, including sensor/actuator faults, software bugs and communication faults. In the present study, the focus has been on investigating a range of sensor-related faults in ASSs. As previously stated, three fault attributes are required as input for the FI test, i.e., location, type and time. The output of LLMs is a json object which consists of a a key-value pair representing the component in which the fault should be introduced by setting the value to "1". The fault TCs output is represented follows:

\[
\mathbf{x} = \{(s,v) \mid s \in S \wedge v \in \{0, 1\}\}
\]

where:

\begin{itemize}
    \item $x$ represents the output of the LLM.
    \item $S$ represents the sensors related to to the fault locations.
    \item $v = 1$ indicates that the fault should be injected in the $s$ location, e.g., specific sensor or actuator.
\end{itemize}

It is noteworthy that the sum of all values in the key-value is typically equivalent to 1 or 2, provided that the faults are injected into one component or two concurrent components at a time according to ISO 26262. 
For examples, considering the aforementioned FSRs, the output of the LLMs are presented as the following:
\begin{itemize}
    \item FSR1: sensor-related faults,
\end{itemize}
    \begin{verbatim}
    { "s1": 1, "s2": 0, "s3": 0,
      "s4": 0, "s5": 0 }
    \end{verbatim}

This key-value indicates that there are 5 potential sensors, and the fault is injected into one location, i.e., vehicle speed sensor.
\begin{itemize}
    \item FSR2: actuator-related faults,
\end{itemize}
    \begin{verbatim}
    { "a1": 0, "a2": 1, "a3": 1 }
    \end{verbatim}
This vector indicates that there are 3 potential actuators, and the fault is injected into two actuators simultaneously, i.e., throttle and brake actuators. Thus, the proposed approach is capable of generating not only a single fault, but also multiple fault locations in the event of a simultaneous faults occurrence. The time at which the fault occurs is determined by the test engineer based on the driving cycle. Notably, integrating an automatic method, e.g., extracting injection points from typical driving profiles and using lightweight heuristics based on velocity and acceleration, is planed to incorporate it in future investigation. This study considers various types of faults related to sensor and actuator signals. Specifically, the sensor-related fault types encompass gain, offset, stuck-at, delay, noise, packet loss, drift and spike. On the other hand, actuator related faults are defined as stuck-at, delay and packet loss faults. To improve the credibility and comparability, the standard CAN fault sets have been considered in this study, i.e., timing-related faults, data-related faults and physical layer faults. The mathematical equations for determining gain, offset, stuck-at, delay, noise, packet loss, drift and spike are demonstrated in the Equation \ref{equ:gain} - \ref{equ:spike}, respectively. 

\begin{equation}
f(t) = g \cdot h(t)
\label{equ:gain}
\end{equation}

\begin{equation}
f(t) =   h(t) + b 
\label{equ:offset}
\end{equation}

\begin{equation}
f(t) =  b 
\label{equ:stuck-at}
\end{equation}

\begin{equation}
 f(t) =  h(t- \tau)
\label{equ:delay}
\end{equation}

\begin{equation}
f(t) =   h(t) +  n(t)
\label{equ:noise}
\end{equation}

\begin{equation}
f(t) = 
\begin{cases} 
x(t), & \text{with probability } p \\
0 \text{ or undefined}, & \text{with probability } 1-p
\end{cases}
\label{equ:packet loss}
\end{equation}

\begin{equation}
f(t) =   h(t) + d(t)
\label{equ:drift}
\end{equation}

\begin{equation}
f(t) =   h(t) + s(t)
\label{equ:spike}
\end{equation}

where \( f(t)\) represents the faulty signal, \(h(t)\) is the healthy or fault-free signal,  \(g\) and \(b\) represent a again and offset value, respectively. \( n(t)\), \( d(t)\) and \( s(t)\) represent the noise, drift and spike functions, respectively.

Once the FI parameters have been identified and transmitted to the fault injector, the execution phase of the fault TCs is initiated. 

\begin{figure}[thb]

    \centering
	\includegraphics[trim={0.1cm 0.1cm 0.1cm 0.1cm}, clip,width=1\linewidth]{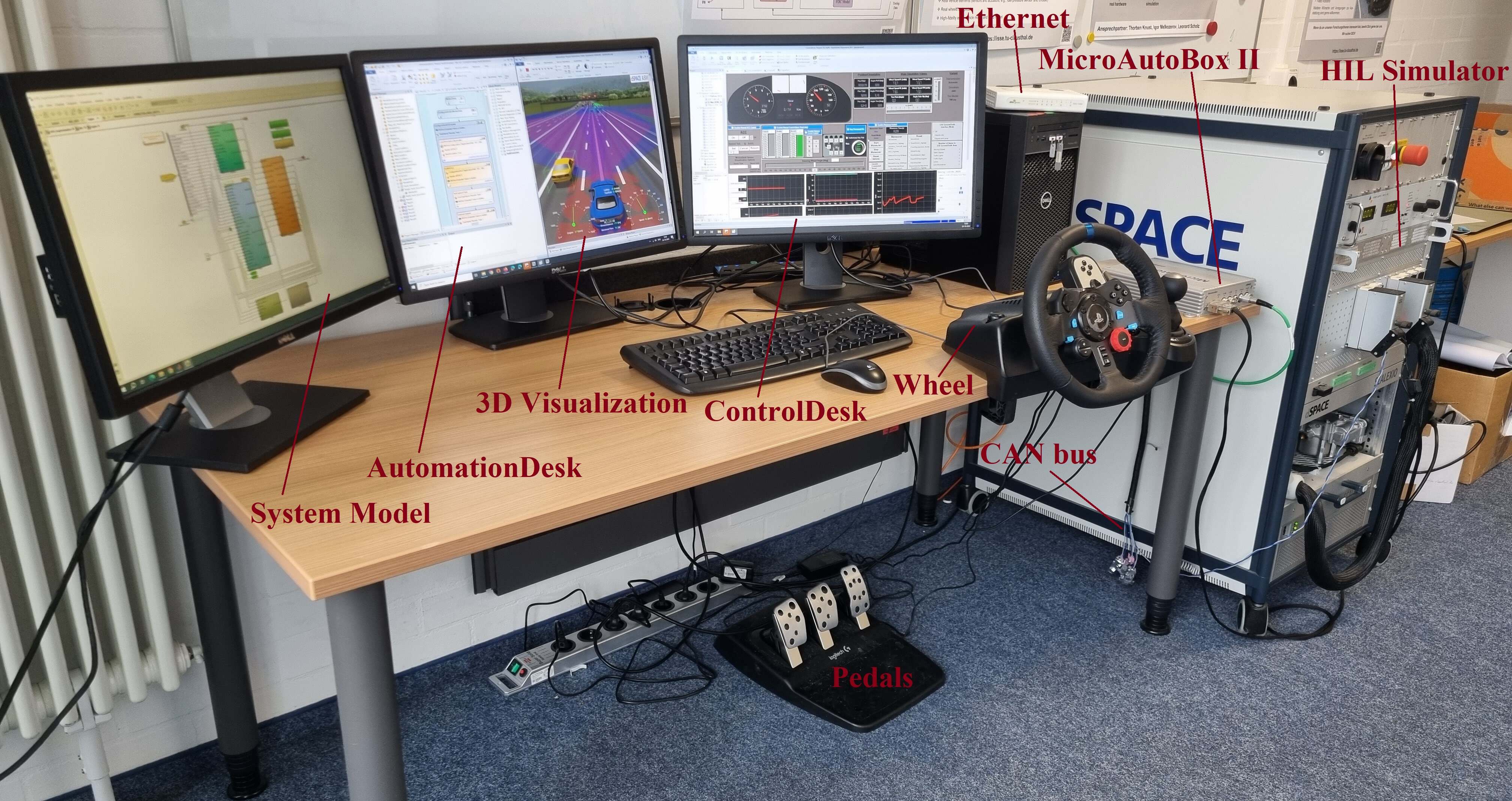}
	\caption{Real-time Hardware-in-the-Loop simulation system.}
	\label{fig:HIL}       
\end{figure}

In order to facilitate the implementation of the FI process during the real-time validation of ASSs, real-time FI framework and HIL simulation platform, developed in \cite{abboush2024representative}, is employed in this study. Through the FI framework, CAN signals are accessed and processed in real time according to fault configurations. By doing so, the system architecture can be ensured as black-box without modification during the FI process. More information can be found in \cite{abboush2022hardware} The principal components of the HIL system are the HIL simulator, actual ECU, CAN bus, physical steering wheel and pedals, and software analysis, modelling and visualisation tools. An illustration of these components is provided in Figure \ref{fig:HIL}. By means of the real-time interface CAN multimessage blockset (RTICANMM), it is possible to access all signals during the real-time execution of the system model on the target machine. Consequently, the aforementioned interface allows for the FI process to be conducted, thereby ensuring the structural integrity of the system model remains unaltered. In other words, the proposed approach guarantees performing the FI test on the SUT as a black box, obviating the necessity for an additional block to simulate the failure mode. The FI test yields data on the reaction of the SUT in a faulty condition, taking into account the surrounding environment.

The subsequent phase is the analysis of the test results, which should occur following the capture of both normal and faulty system behavior. The normal behavior is recorded by executing the system model code in the HIL under fault-free conditions, which is called 'golden run'. This behavior serves as a reference for the subsequent failure analysis process. Conversely, the malfunctioning behavior is captured by injecting the fault TCs generated by LLMs at the component level and monitoring the system variables at the system level. 

For illustration, according to the above FSRs, a gain fault is injected into the vehicle speed sensor and the system variables such as engine speed, torque, temperature, and vehicle dynamics are recorded. By comparing the faulty and normal behavior, the robustness of the SUT and the safety mechanisms can be evaluated. In other words, it can be determined whether the FSRs are met by the behavior of the SUT. In the event of a violation against the FSRs, a feedback report is created by the test engineers and sent to the development phase, indicating the necessary measures.

\section{Experiments and Results} \label{sec:expr}

This section outlines the experimental setup and results of our LLM-driven approach for real-time fault injection, which generates fault test cases (TCs) from natural language functional safety requirements (FSRs). The evaluation proceeds in two stages: first, classifying each FSR as sensor- or actuator-related; second, generating fault TCs based on the classification and a runtime-provided list of supported sensors. All TCs comply with ISO~26262, limiting fault models to two concurrent faults to meet safety-critical constraints.

Experiments used a curated FSR dataset from prior work \cite{amyan2024automating}. To ensure adaptability to dynamic automotive environments, we applied few-shot prompting \cite{fewshots}, enabling real-time injection of contextual knowledge—such as updating supported sensors—without retraining. Unlike fine-tuning, this method maintains flexibility across domain shifts without incurring heavy computational costs.

All models were prompted using reusable templates available in our repository \cite{repo}. We tested both proprietary (OpenAI’s \texttt{gpt-4o 2024-11-20} and \texttt{gpt-4o-mini 2024-07-18} \cite{gpt4o, gpt4o_mini}) and open-source models (\texttt{phi-4-14b} via Ollama \cite{phi4, ollama}, and \texttt{qwen2.5-7b-instruct}, \texttt{llama-3-70b-instruct} via Novita AI \cite{qwen2, llama3modelcard}), all evaluated under a uniform configuration: temperature = 0.2, seed = 42.

We used accuracy and F1-macro to assess performance. While accuracy measures overall correctness, the F1-macro—our primary metric—averages class-wise F1 scores, offering a fair evaluation despite the sensor-actuator class imbalance.

\subsection{Classifing Actuator and Sensor-Based FSRs} \label{exp1}

The first experimental stage focused on categorizing each FSR as either actuator-related or sensor-related.
The dataset curated from prior work \cite{amyan2024automating} and consists of 134 total FSRs: 97 related to sensors and 37 to actuators.
Table~\ref{tab:actuator_stats} summarizes the F1 scores obtained by each LLM for this binary classification task.
The best-performing model was \texttt{gpt-4o} with $N=1$, achieving an F1 score of \textit{88.0\%}. This was followed closely by \texttt{qwen2.5-7b} ($N=5$, \textit{85.7\%}), \texttt{phi-4-14b} ($N=5$, \textit{83.2\%}), \texttt{llama-3-70b} ($N=3$, \textit{82.4\%}), and \texttt{gpt-4o-mini} ($N=1$, \textit{75.4\%}). These results confirm the suitability of LLMs for accurate classification of natural language FSRs.

\begin{table}[h!]
    \centering
    \caption{FSRs Classification Performance of various LLMs (\%)}
    \label{tab:actuator_stats}
    \begin{tabular}{l cc | cc | cc}
        \toprule
        \multirow{2}{*}{Model} & \multicolumn{6}{c}{Number of Examples} \\
        \cmidrule(lr){2-7}
         & \multicolumn{2}{c}{1} & \multicolumn{2}{c}{3} & \multicolumn{2}{c}{5} \\
         \cmidrule(lr){2-3}\cmidrule(lr){4-5}\cmidrule(lr){6-7}
         & Acc & F1 & Acc & F1 & Acc & F1 \\
        \midrule
        gpt-4o & 90.3 & \textbf{88.0} & 86.6 & 84.0 & 85.1 & 82.7 \\
        gpt-4o-mini & 77.6 & \textbf{75.4} & 68.7 & 67.4 & 70.1 & 68.3 \\
        llama-3-70b & 86.6 & 80.2 & 85.8 & \textbf{82.4} & 79.9 & 77.5 \\
        phi4-14b & 70.9 & 69.5 & 85.1 & 82.4 & 85.8 & \textbf{83.2} \\
        qwen2.5-7b & 85.8 & 82.1 & 84.3 & 81.2 & 88.1 & \textbf{85.7} \\
        \bottomrule
    \end{tabular}
\end{table}

\subsection{Fault TCs generation based on classified FSRs}  \label{exp2}

 The second stage evaluates LLMs’ ability to generate fault test cases from FSRs classified as sensor-related, under ISO~26262 constraints allowing a maximum of two concurrent faults. Accordingly, the dataset reflects this limit, with each FSR involving up to two simultaneous sensor faults.

To guide generation within system capabilities, a predefined list of supported sensors—APP, WSA, WS, YR, and ST—was dynamically included in the prompt, along with short functional descriptions for context. This design leverages few-shot prompting to enable real-time reconfiguration without retraining, underscoring its flexibility over fine-tuning.

\begin{figure}[htbp]
\centering
\includegraphics[scale=0.6]{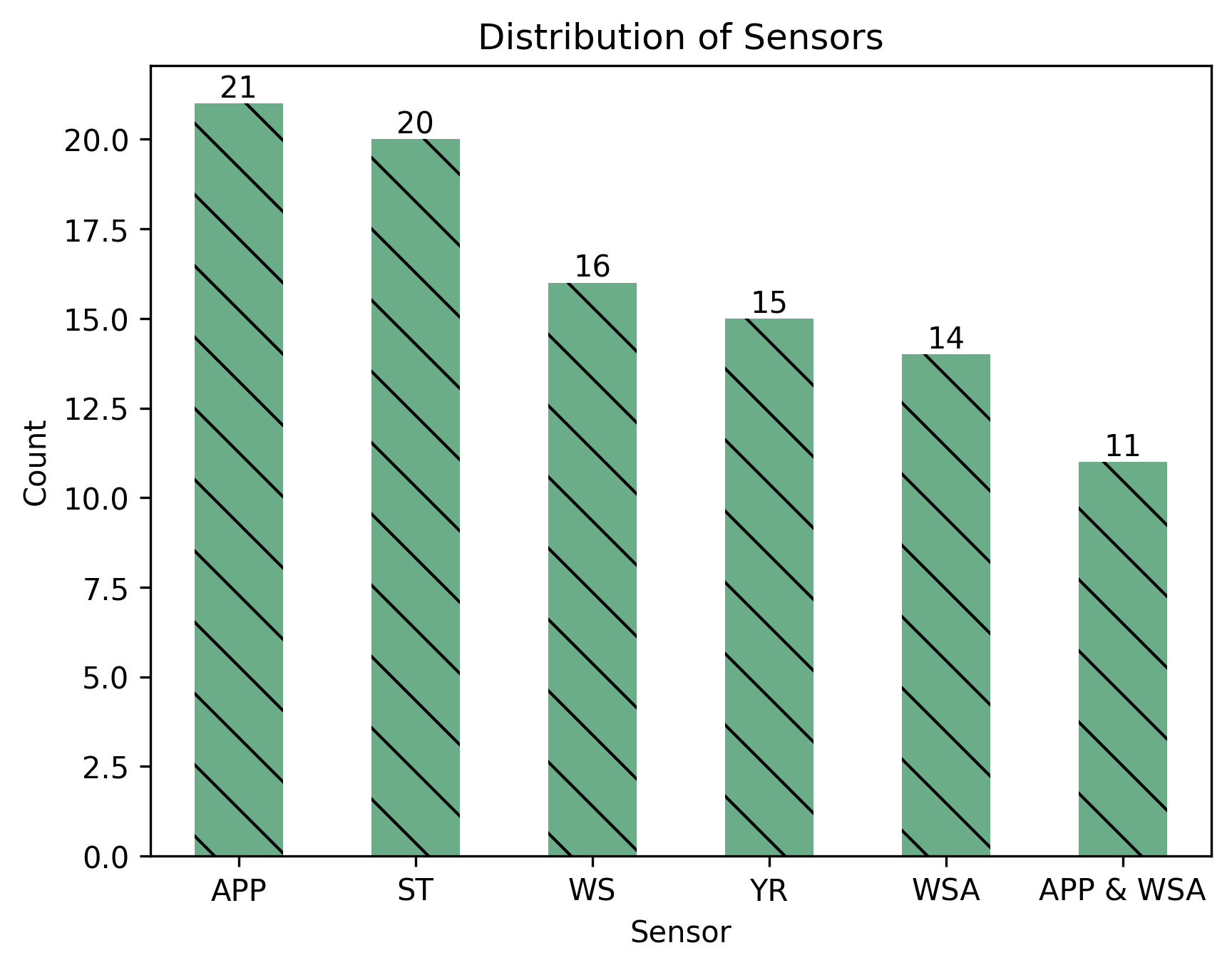}
\caption{Number of FSR per Sensor.}
\label{fig:num_per_sensor}
\end{figure}

We conducted three trials to assess generation effectiveness: (1) single TC per FSR, (2) batch generation in a single prompt, and (3) testing model behavior when relevant sensors are excluded. Figure~\ref{fig:num_per_sensor} illustrates the FSR distribution across sensors, highlighting class representation.

\subsubsection{Single Fault Test Case} \label{exp_trial_1}

In this trial, one fault TC was generated per request, each based on a single FSR. Table~\ref{tab:single_stats} presents the performance of each model. The highest performance was achieved by \texttt{gpt-4o} at $N=8$ with an F1 score of \textit{97.5\%}. Notably, the same model reached \textit{97.0\%} with $N=1$, reducing token usage while maintaining performance.

Among open-source models, \texttt{phi-4-14b} achieved consistent F1 scores across all $N$ values. This consistency was verified through repeated trials on the locally hosted Ollama server. However, the model failed when provided with a large number of examples per request. In these cases, the generated outputs deviated from the expected format, resulting in an F1 score of \textit{0\%}. This failure is attributed to the model exceeding its context window, which caused the format instructions in the prompt to be ignored and led to unstructured output. Other noteworthy performances include \texttt{phi-4-14b} ($N=3$, \textit{96.6\%}), \texttt{llama-3-70b} ($N=1$, \textit{95.2\%}), \texttt{gpt-4o-mini} ($N=1$, \textit{91.0\%}), and \texttt{qwen2.5-7b} ($N=8$, \textit{89.9\%}).

\begin{table}[h!]
    \centering
    \caption{Single Sensor-Based Model Accuracy (\%)}
    \label{tab:single_stats}
    \begin{tabular}{lcc|cc|cc|cc}
        \toprule
        \multirow{2}{*}{Model} & \multicolumn{8}{c}{Number of Examples} \\
        \cmidrule(lr){2-9}
         & \multicolumn{2}{c}{1} & \multicolumn{2}{c}{3} & \multicolumn{2}{c}{5} & \multicolumn{2}{c}{8} \\
         \cmidrule(lr){2-3}\cmidrule(lr){4-5}\cmidrule(lr){6-7}\cmidrule(lr){8-9}
         & Acc & F1 & Acc & F1 & Acc & F1 & Acc & F1 \\
        \midrule
        gpt-4o & 94.8 & 97.0 & 94.8 & 95.8 & 95.9 & 96.7 & 95.9 & \textbf{97.5} \\
        gpt-4o-mini & 81.4 & \textbf{91.0} & 78.4 & 87.8 & 82.5 & 89.7 & 80.4 & 89.2 \\
        llama-3-70b & 91.8 & \textbf{95.2} & 90.7 & 95.1 & 86.6 & 92.8 & 90.7 & 94.6 \\
        phi4-14b & 93.8 & 96.1 & 93.8 & \textbf{96.6} & 93.8 & 96.6 & 0.0 & 0.0 \\
        qwen2.5-7b & 80.4 & 86.7 & 75.3 & 86.1 & 79.4 & 88.8 & 83.5 & \textbf{89.9} \\
        \bottomrule
    \end{tabular}
\end{table}

\subsubsection{Batch of Test Cases} \label{exp_trial_2}

This trial tested the efficiency of generating multiple fault TCs in a single batch request, which is valuable for real-world scalability. The experiment varied the number of examples ($N$) and batch size (BS), i.e., number of FSRs per batch. The results are presented in Table~\ref{tab:batch_size_grouped}.

Preference was given to configurations with lower token usage when the F1 scores were identical. For example, \texttt{gpt-4o} yielded equal F1 scores with $(\textit{BS=3}, \textit{N=1}, \textit{tokens}=143{,}952)$ and $(\textit{BS=5}, \textit{N=8}, \textit{tokens}=303{,}310)$. The former configuration was favored for its efficiency.

Optimal configurations include: \texttt{llama-3-70b} ($\textit{BS=2}, \textit{N=3}$), \texttt{phi-4-14b} ($\textit{BS=5}, \textit{N=1}$), \texttt{gpt-4o-mini} ($\textit{BS=3}, \textit{N=8}$), and \texttt{qwen2.5-7b} ($\textit{BS=2}, \textit{N=3}$). These combinations demonstrated strong performance while balancing processing efficiency.

\begin{table}[h!]
    \caption{Model Performance Grouped by Batch Size (\%)}
    \label{tab:batch_size_grouped}
    \setlength{\tabcolsep}{4pt}
    \begin{tabular}{llcc|cc|cc|cc}
        \toprule
        \multirow{2}{*}{\shortstack[c]{BS}} & \multirow{2}{*}{Model} & \multicolumn{8}{c}{Number of Examples} \\
        \cmidrule(lr){3-10}
         & & \multicolumn{2}{c}{1} & \multicolumn{2}{c}{3} & \multicolumn{2}{c}{5} & \multicolumn{2}{c}{8} \\
         \cmidrule(lr){3-4}\cmidrule(lr){5-6}\cmidrule(lr){7-8}\cmidrule(lr){9-10}
         & & Acc & F1 & Acc & F1 & Acc & F1 & Acc & F1 \\
        \midrule
        \multirow{5}{*}{2} & gpt-4o & 92.7 & 95.4 & 93.8 & 95.6 & 95.8 & 97.2 & 93.8 & 95.2 \\
         & gpt-4o-mini & 87.5 & 92.7 & 87.5 & 94.1 & 87.5 & 92.3 & 89.6 & 93.7 \\
         & llama-3-70b & 92.7 & 96.4 & 94.8 & \textbf{97.3} & 91.7 & 94.8 & 93.8 & 96.8 \\
         & phi4-14b & 93.8 & 96.5 & 92.7 & 95.1 & 0.0 & 0.0 & 0.0 & 0.0 \\
         & qwen2.5-7b & 76.0 & 85.4 & 83.3 & \textbf{88.9} & 77.1 & 87.2 & 81.2 & 86.7 \\
        \midrule
        \multirow{5}{*}{3} & gpt-4o & 95.8 & \textbf{97.5} & 95.8 & 97.0 & 95.8 & 97.2 & 94.8 & 96.4 \\
         & gpt-4o-mini & 85.4 & 91.9 & 83.3 & 91.1 & 91.7 & 93.8 & 90.6 & \textbf{94.1} \\
         & llama-3-70b & 91.7 & 94.9 & 88.5 & 93.8 & 92.7 & 95.7 & 92.7 & 95.5 \\
         & phi4-14b & 95.8 & 96.4 & 95.8 & 96.4 & 0.0 & 0.0 & 0.0 & 0.0 \\
         & qwen2.5-7b & 78.1 & 86.0 & 81.2 & 85.8 & 83.3 & 88.1 & 77.1 & 84.7 \\
        \midrule
        \multirow{5}{*}{5} & gpt-4o & 92.6 & 95.4 & 93.7 & 96.5 & 94.7 & 96.6 & 95.8 & 97.5 \\
         & gpt-4o-mini & 87.4 & 93.6 & 86.3 & 93.4 & 87.4 & 92.1 & 87.4 & 92.3 \\
         & llama-3-70b & 91.6 & 95.1 & 93.7 & 96.1 & 88.4 & 94.4 & 91.6 & 95.5 \\
         & phi4-14b & 95.8 & \textbf{96.5} & 93.7 & 94.9 & 0.0 & 0.0 & 0.0 & 0.0 \\
         & qwen2.5-7b & 76.8 & 84.4 & 77.9 & 84.5 & 80.0 & 87.2 & 78.9 & 85.9 \\
        \bottomrule
    \end{tabular}
\end{table}

\subsubsection{Dropping Sensors} \label{exp_trial_3}

In the final trial, we assessed whether models could correctly avoid generating fault TCs for unsupported sensors. Two sensors—Wheel Steering Angle (WSA) and Steering Torque (ST)—were deliberately omitted from the sensor list passed in the prompt. Two FSRs related to these sensors were used as negative examples, and the expected output was an empty list.

Table~\ref{tab:dropped_stats} presents the F1 scores from this trial. The top performer was \texttt{phi-4-14b} ($N=5$, \textit{97.7\%}), followed by \texttt{gpt-4o-mini} ($N=3$, \textit{96.8\%}), \texttt{gpt-4o} ($N=5$, \textit{96.1\%}), \texttt{qwen2.5-7b} ($N=8$, \textit{92.7\%}), and \texttt{llama-3-70b} ($N=1$, \textit{92.6\%}). These results indicate the models' ability to filter out unsupported sensors, a critical feature for robust TC generation.

\begin{table}[h!]
\caption{Dropped Sensor (Single) Model Accuracy (\%)}
\label{tab:dropped_stats}
    \centering
    \begin{adjustbox}{width=\columnwidth,center}
    \begin{tabular}{l cc | cc | cc | cc}
        \toprule
        \multirow{2}{*}{Model} & \multicolumn{8}{c}{Number of Examples} \\
        \cmidrule(lr){2-9}
         & \multicolumn{2}{c}{1} & \multicolumn{2}{c}{3} & \multicolumn{2}{c}{5} & \multicolumn{2}{c}{8} \\
         \cmidrule(lr){2-3}\cmidrule(lr){4-5}\cmidrule(lr){6-7}\cmidrule(lr){8-9}
         & Acc & F1 & Acc & F1 & Acc & F1 & Acc & F1 \\
        \midrule
        gpt-4o & 93.8 & 93.9 & 91.8 & 92.5 & 94.8 & \textbf{96.1} & 92.8 & 94.7 \\
        gpt-4o-mini & 93.8 & 94.5 & 95.9 & \textbf{96.8} & 92.8 & 94.5 & 89.7 & 92.2 \\
        llama-3-70b & 91.8 & \textbf{92.6} & 83.5 & 86.3 & 86.6 & 90.6 & 85.6 & 88.7 \\
        phi4-14b & 93.8 & 96.4 & 92.8 & 96.8 & 94.8 & \textbf{97.7} & 0.0 & 0.0 \\
        qwen2.5-7b & 85.6 & 86.7 & 83.5 & 86.0 & 88.7 & 91.2 & 90.7 & \textbf{92.7} \\
        \bottomrule
    \end{tabular}
    \end{adjustbox}
\end{table}

\subsection{LLMs Result's Discussion}

Table~\ref{tab:actuator_stats} shows that classifying FSRs as actuator- or sensor-related remains a non-trivial task for LLMs. \texttt{gpt-4o} achieved the highest F1-macro at \textit{88.0\%}, demonstrating strong generalization even with minimal in-context examples. In contrast, \texttt{gpt-4o-mini} scored lowest at \textit{75.4\%}, highlighting how limited model capacity can constrain performance. These results point to the potential of advanced prompting strategies—like Chain-of-Thought \cite{cot} and Feedback-based prompting \cite{feedback_loop}—to improve weaker models.

Notably, \texttt{qwen2.5-7b} surpassed \texttt{gpt-4o-mini} when given more examples, suggesting that smaller models benefit more from richer contextual input, underscoring the need to adapt prompt size to model capability.

\texttt{phi-4-14b} showed a split profile: modest classification performance in \ref{exp1}, yet near-optimal F1 scores in all fault TC generation trials (\ref{exp2}), trailing \texttt{gpt-4o} by no more than 1\%. Its exclusive training on high-quality synthetic data \cite{phi4} likely enhances its robustness in structured generation—consistent with its strong showing on reasoning benchmarks like GPQA \cite{gpqa}, where it outperforms larger proprietary models.

\texttt{Llama-3-70b} delivered stable results, particularly with fewer examples. In contrast, \texttt{qwen2.5-7b} required more extensive prompting, revealing a sharper sensitivity to input richness that may hinder real-time use under tight latency or input constraints.

In sum, \texttt{gpt-4o} remains the top closed-source choice for high-performance use where infrastructure is not a bottleneck, while \texttt{phi-4-14b} emerges as a strong open-source alternative—combining competitive accuracy with local deployment availability, ideal for applications prioritizing offline access, data sovereignty, or cost control.

\subsection{Fault TCs Execution and Effect Analysis}

\begin{figure}[thb]

    \centering
	\includegraphics[trim={0.1cm 0.1cm 0.1cm 0.1cm}, clip,width=1\linewidth]{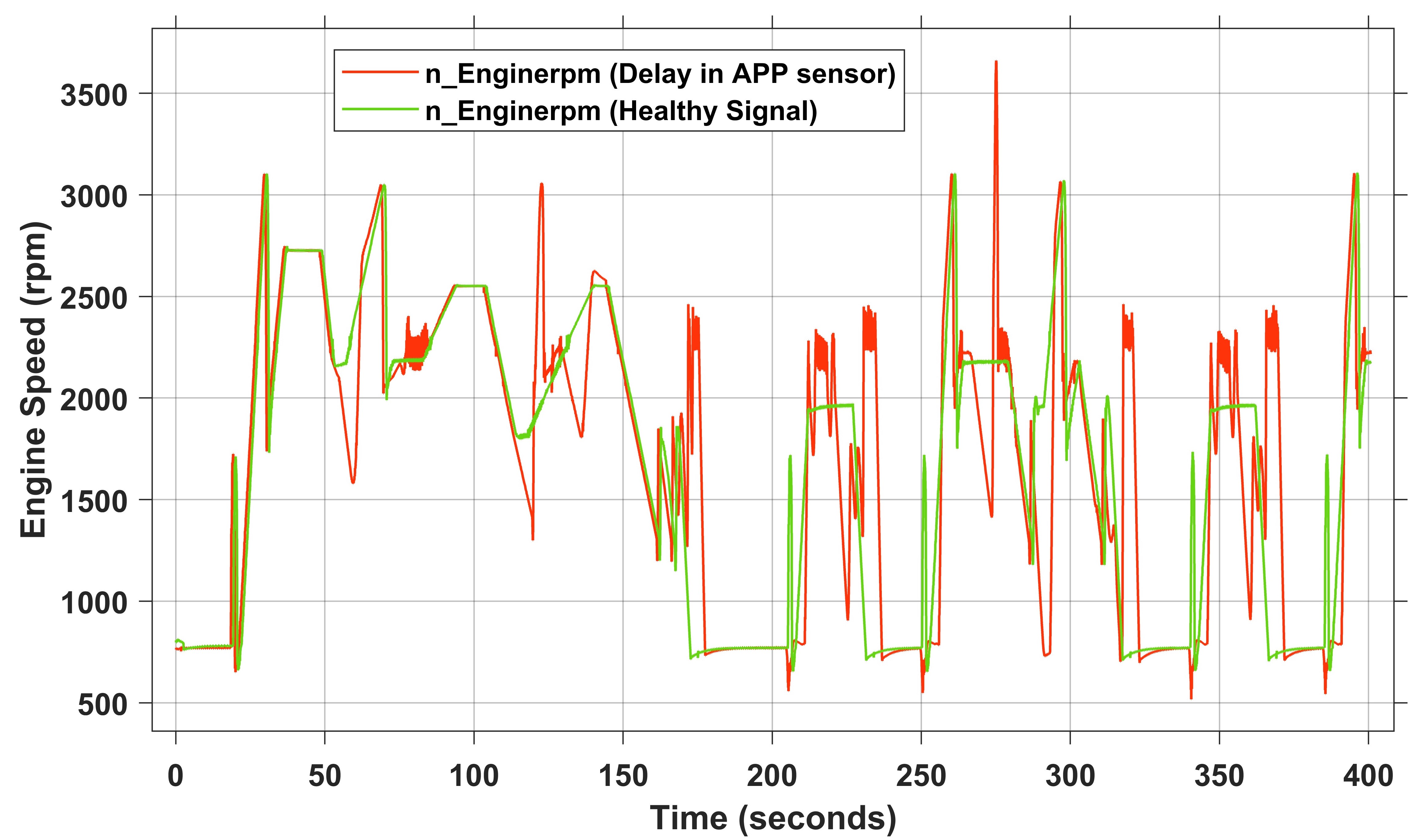}
	\caption{Fault effect analysis during the delay fault occurrence in \textit{APP} sensor.}
	\label{fig:fault}       
\end{figure}

After LLMs generate the fault TCs, corresponding fault locations are identified and injected in real-time via RTICANMM within the ASS simulation environment. To isolate critical faults that violate FSRs, system behavior under faulty conditions is compared to the fault-free baseline. As shown in Figure~\ref{fig:fault}, injecting a delay fault into the \textit{APP} sensor caused unacceptable engine speed deviations between 175–375 seconds, leading to FSR violations during driving. However, within [0–120] seconds, the system successfully mitigated the fault, maintaining stable operation.

\begin{figure}[thb]

    \centering
	\includegraphics[trim={0.1cm 0.1cm 0.1cm 0.1cm}, clip,width=0.9\linewidth]{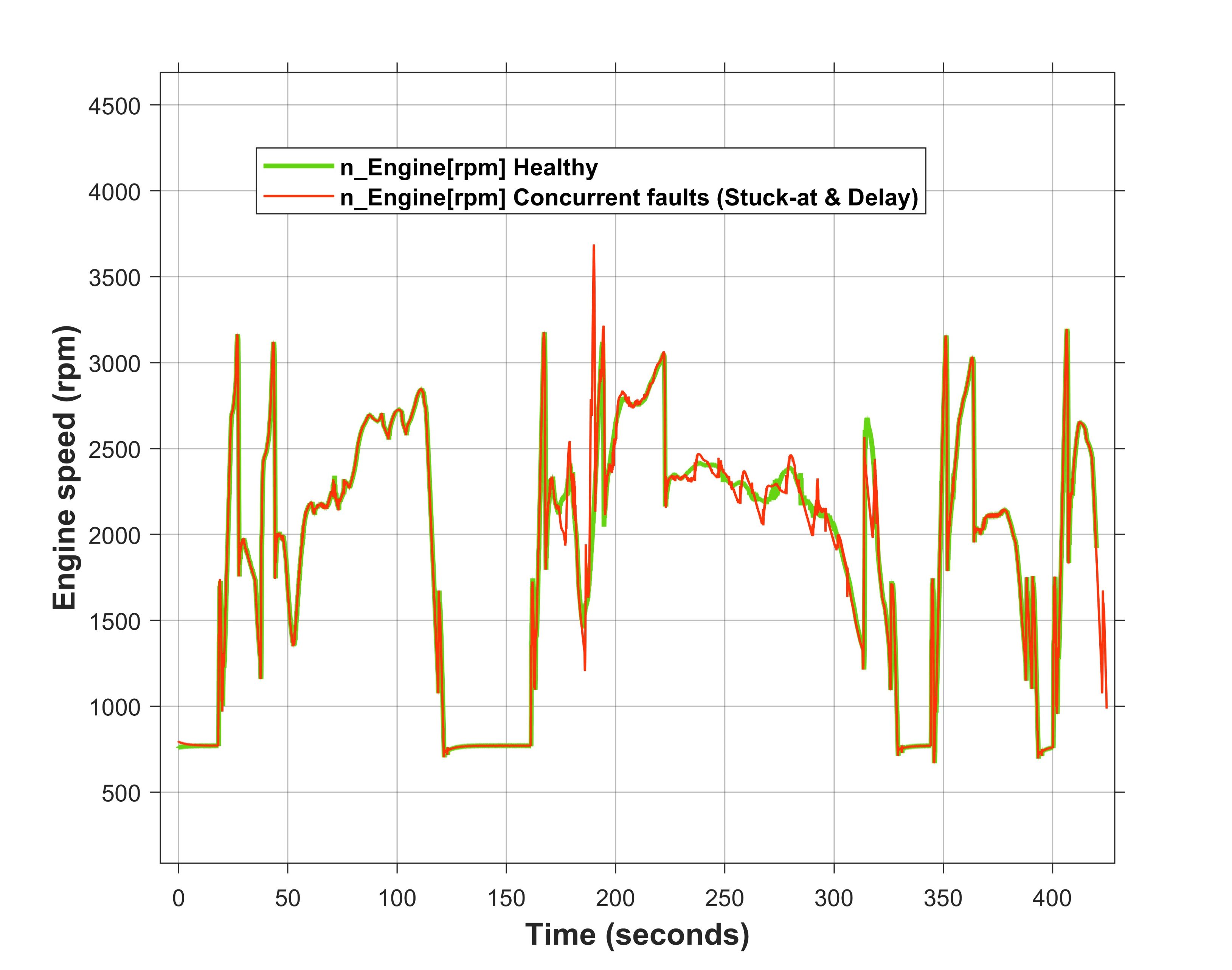}
	\caption{Engine speed behavior under concurrent faults.}
	\label{fig:concurrent fault1}       
\end{figure}

\begin{figure}[thb]

    \centering
	\includegraphics[trim={0.1cm 0.1cm 0.1cm 0.1cm}, clip,width=0.9\linewidth]{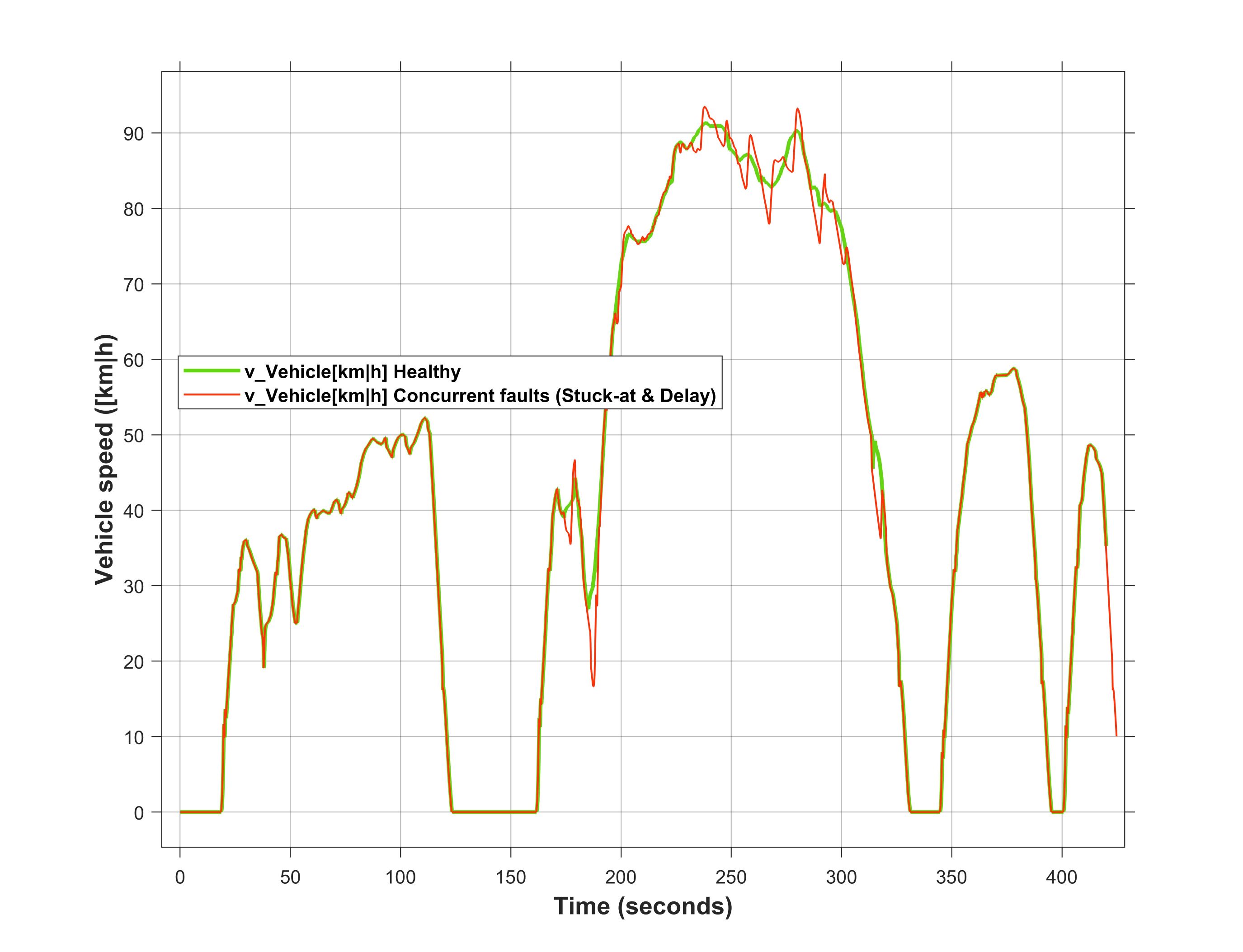}
	\caption{ Vehicle speed behavior under concurrent faults.}
	\label{fig:concurrent fault2}       
\end{figure}

\begin{figure}[thb]

    \centering
	\includegraphics[trim={0.1cm 0.1cm 0.1cm 0.1cm}, clip,width=0.9\linewidth]{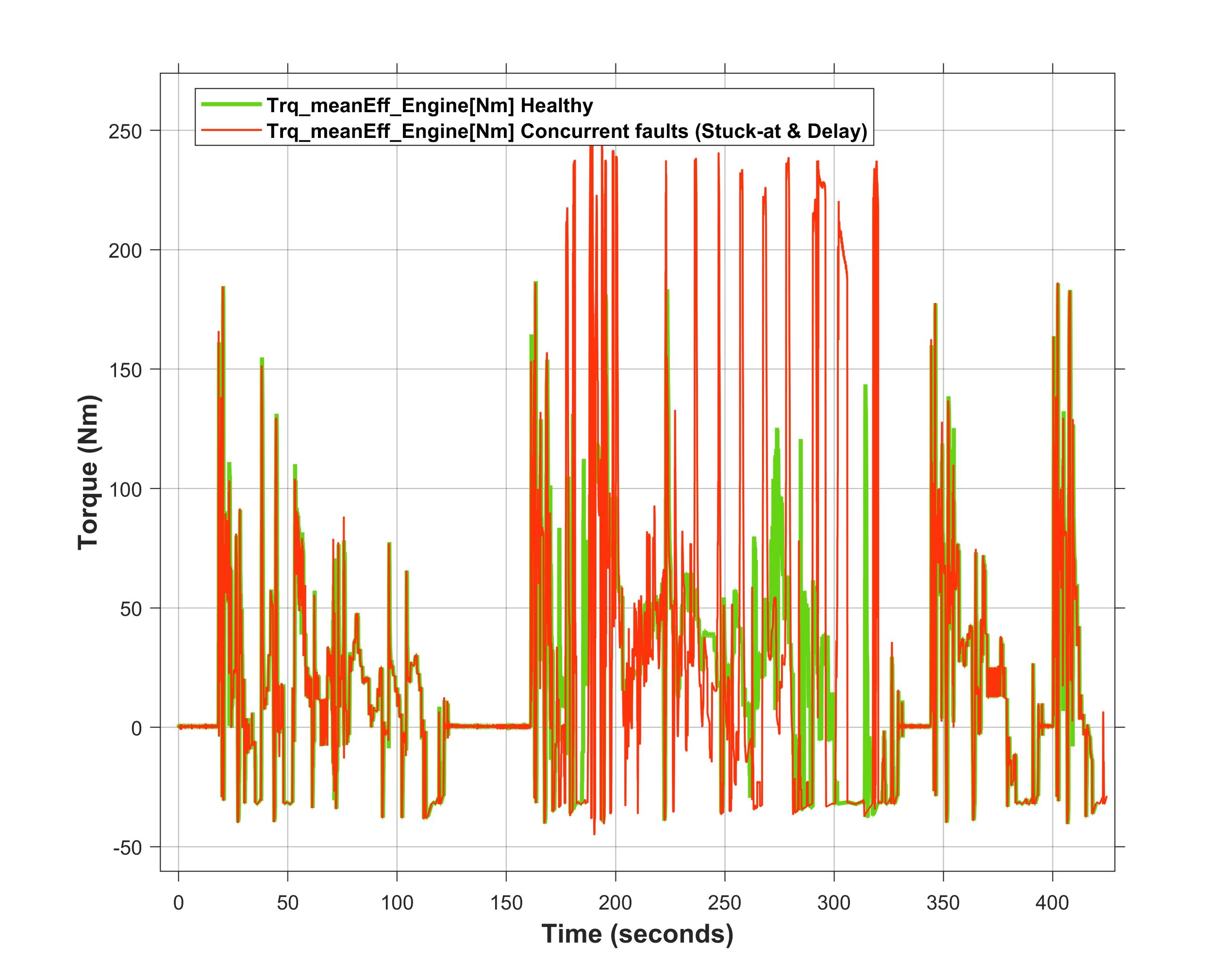}
	\caption{Fault effect on vehicle torque concurrent faults.}
	\label{fig:concurrent fault3}       
\end{figure}

\begin{figure}[thb]

    \centering
	\includegraphics[trim={0.1cm 0.1cm 0.1cm 0.1cm}, clip,width=0.9\linewidth]{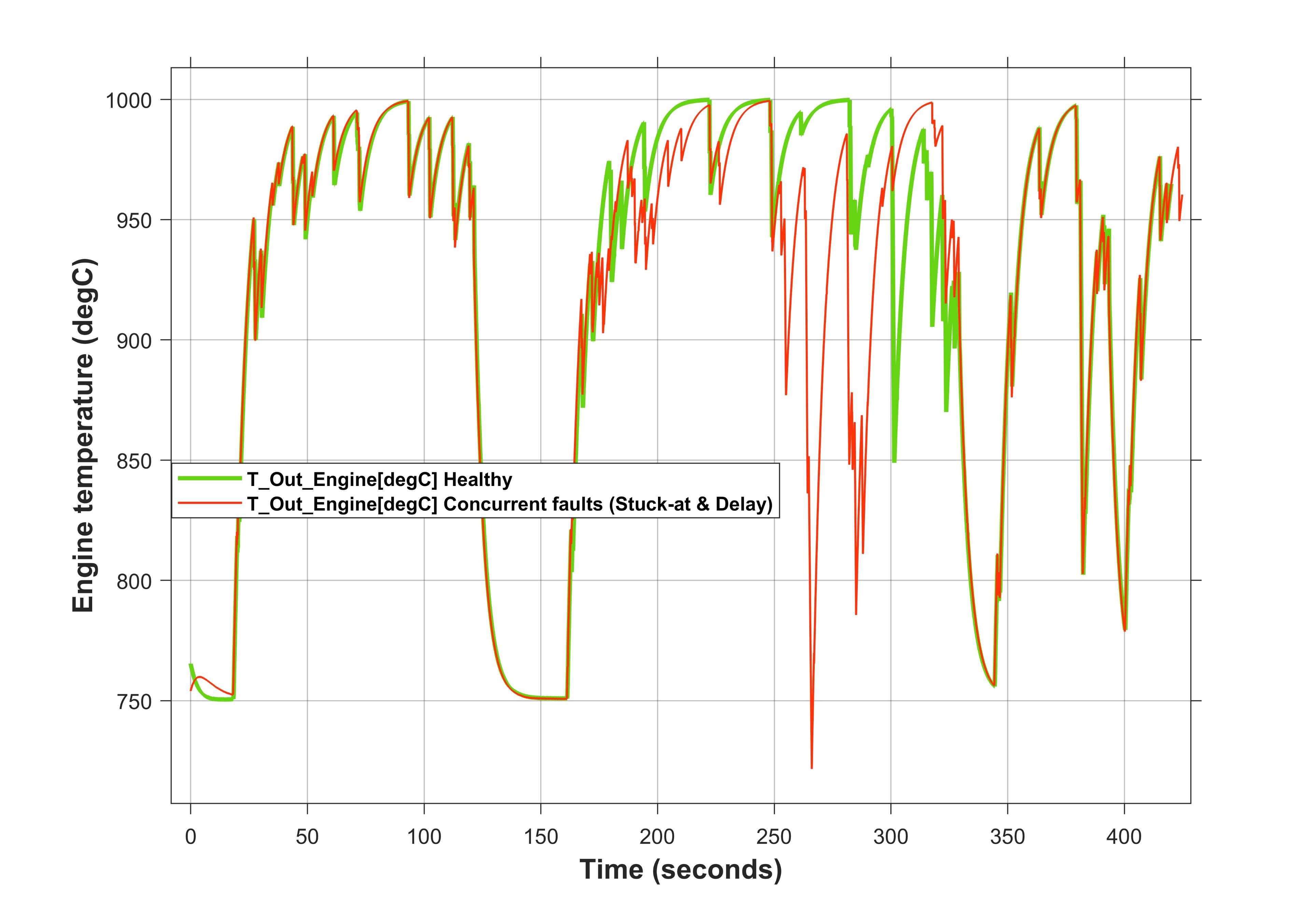}
	\caption{Engine temperature behavior under concurrent faults.}
	\label{fig:concurrent fault4}       
\end{figure}

To assess the impact of concurrent faults, two distinct fault types were injected at separate locations, based on their associated FSRs. Figure~\ref{fig:concurrent fault1} illustrates a delay fault in the RPM sensor combined with a stuck-at fault in the APP sensor, revealing a system transition from safe to faulty behavior. In Figure~\ref{fig:concurrent fault2}, the vehicle fails to behave as intended between 170–300 seconds, although recovery is briefly observed between [190–230] seconds where vehicle speed stabilizes. Figure~\ref{fig:concurrent fault3} shows high torque applied to compensate for abnormal sensor amplitudes, leading to increased energy demand. As depicted in Figure~\ref{fig:concurrent fault4}, this ultimately destabilizes engine temperature, with a notable drop between [250–290] seconds, resulting in anomalous system-level behavior and FSR violations.

From these analyses, the system state under critical fault conditions—including fault type, value, and timing—was extracted during runtime. These results are compiled into a test report and forwarded to the development team to support corrective actions within the target system.


\section{Conclusion} \label{sec:conc}
This paper presents a novel approach enabling the intelligent generation of fault TCs for real-time FI testing of ASSs on a HIL
system. The primary objective is to automatically and systematically generate representative fault TCs for the purpose of validating the FSRs during the development phase of ASSs. In this respect, the feasibility of incorporating a variety of LLMs into the real-time FI process has been explored. The experimental results demonstrated that the \texttt{gpt-4o} exhibited a superior performance in fault TCs generation compared to other models, with an F1-score of 97.5\%. Notably, as an open-source LLMs, \texttt{phi4-14b}, have demonstrated efficacy in accurately performing the task of generating fault TCs, as evidenced by a high F1-score of 96.6\%. The execution of the generated TCs demonstrates effectiveness of the proposed approach in analysing system behavior under the generated fault TCs. The proposed approach therefore not only supports safety engineers, but also optimizes the testing process in terms of performance, effort, and costs.

As future work, incorporating other communication-level faults and ECU-level failures is essential to capturing the full spectrum of fault modes encountered in real-world automotive systems. Furthermore, investigating the feasibility of LLMs in providing a rationale behind the prediction outcome to support decision making can be an extension of the proposed work. Few-shot prompting offer a good generalization to other types of requirements making this solution extendable.


\bibliographystyle{IEEEtran}
\bibliography{bibliography}

@inproceedings{abelein2012complexity,
  title={Complexity, quality and robustness-the challenges of tomorrow's automotive electronics},
  author={Abelein, Ulrich and Lochner, Helmut and Hahn, Daniel and Straube, Stefan},
  booktitle={2012 Design, Automation \& Test in Europe Conference \& Exhibition (DATE)},
  pages={870--871},
  year={2012},
  organization={IEEE}
}

@article{garousi2018testing3,
  title={Testing embedded software: A survey of the literature},
  author={Garousi, Vahid and Felderer, Michael and Karap{\i}{\c{c}}ak, {\c{C}}a{\u{g}}r{\i} Murat and Y{\i}lmaz, U{\u{g}}ur},
  journal={Information and Software Technology},
  volume={104},
  pages={14--45},
  year={2018},
  publisher={Elsevier}
}

@misc{url,
author = {},
title = {ISO 26262-1:2018 - Road vehicles — Functional safety — Part 4},
howpublished = {\url{https://www.iso.org/standard/68383.html}},
month = {},
year = {},
note = {(Accessed on 10/22/2024)}
}

@inproceedings{pintard2014safety,
  title={From safety analyses to experimental validation of automotive embedded systems},
  author={Pintard, Ludovic and Fabre, Jean-Charles and Leeman, Michel and Kanoun, Karama and Roy, Matthieu},
  booktitle={2014 IEEE 20th Pacific Rim International Symposium on Dependable Computing},
  pages={125--134},
  year={2014},
  organization={IEEE}
}

@article{hsueh1997fault,
  title={Fault injection techniques and tools},
  author={Hsueh, Mei-Chen and Tsai, Timothy K and Iyer, Ravishankar K},
  journal={Computer},
  volume={30},
  number={4},
  pages={75--82},
  year={1997},
  publisher={IEEE}
}

@inproceedings{ubar2010parallel,
  title={Parallel X-fault simulation with critical path tracing technique},
  author={Ubar, Raimund and Devadze, Sergei and Raik, Jaan and Jutman, Artur},
  booktitle={2010 Design, Automation \& Test in Europe Conference \& Exhibition (DATE 2010)},
  pages={879--884},
  year={2010},
  organization={IEEE}
}

@inproceedings{zheng2009monte,
  title={A Monte Carlo-based control signal generator for single event effetc (SEE) fault injection},
  author={Zheng, Hongchao and Fan, Long and Yue, Suge and Liu, Liquan},
  booktitle={2009 European Conference on Radiation and Its Effects on Components and Systems},
  pages={247--251},
  year={2009},
  organization={IEEE}
}

@book{benso2003fault,
  title={Fault injection techniques and tools for embedded systems reliability evaluation},
  author={Benso, Alfredo and Prinetto, Paolo},
  volume={23},
  year={2003},
  publisher={Springer Science \& Business Media}
}

@article{natella2012fault,
  title={On fault representativeness of software fault injection},
  author={Natella, Roberto and Cotroneo, Domenico and Duraes, Joao A and Madeira, Henrique S},
  journal={IEEE Transactions on Software Engineering},
  volume={39},
  number={1},
  pages={80--96},
  year={2012},
  publisher={IEEE}
}

@article{zhao2023survey,
  title={A survey of large language models},
  author={Zhao, Wayne Xin and Zhou, Kun and Li, Junyi and Tang, Tianyi and Wang, Xiaolei and Hou, Yupeng and Min, Yingqian and Zhang, Beichen and Zhang, Junjie and Dong, Zican and others},
  journal={arXiv preprint arXiv:2303.18223},
  year={2023}
}

@article{hou2024large,
  title={Large language models for software engineering: A systematic literature review},
  author={Hou, Xinyi and Zhao, Yanjie and Liu, Yue and Yang, Zhou and Wang, Kailong and Li, Li and Luo, Xiapu and Lo, David and Grundy, John and Wang, Haoyu},
  journal={ACM Transactions on Software Engineering and Methodology},
  volume={33},
  number={8},
  pages={1--79},
  year={2024},
  publisher={ACM New York, NY}
}

@article{wang2024softwar,
  title={Software testing with large language models: Survey, landscape, and vision},
  author={Wang, Junjie and Huang, Yuchao and Chen, Chunyang and Liu, Zhe and Wang, Song and Wang, Qing},
  journal={IEEE Transactions on Software Engineering},
  year={2024},
  publisher={IEEE}
}

@article{abboush2022hardware,
  title={Hardware-in-the-loop-based real-time fault injection framework for dynamic behavior analysis of automotive software systems},
  author={Abboush, Mohammad and Bamal, Daniel and Knieke, Christoph and Rausch, Andreas},
  journal={Sensors},
  volume={22},
  number={4},
  pages={1360},
  year={2022},
  publisher={MDPI}
}

@inproceedings{vedder2013combining,
  title={Combining fault-injection with property-based testing},
  author={Vedder, Benjamin and Arts, Thomas and Vinter, Jonny and Jonsson, Magnus},
  booktitle={Proceedings of International Workshop on Engineering Simulations for Cyber-Physical Systems},
  pages={1--8},
  year={2013}
}

@inproceedings{cong2015automatic,
  title={Automatic fault injection for driver robustness testing},
  author={Cong, Kai and Lei, Li and Yang, Zhenkun and Xie, Fei},
  booktitle={Proceedings of the 2015 International Symposium on Software Testing and Analysis},
  pages={361--372},
  year={2015}
}

@inproceedings{khosrowjerdi2018virtualized,
  title={Virtualized-fault injection testing: A machine learning approach},
  author={Khosrowjerdi, Hojat and Meinke, Karl and Rasmusson, Andreas},
  booktitle={2018 IEEE 11th International Conference on Software Testing, Verification and Validation (ICST)},
  pages={297--308},
  year={2018},
  organization={IEEE}
}

@inproceedings{sedaghatbaf2022delfase,
  title={DELFASE: A Deep Learning Method for Fault Space Exploration},
  author={Sedaghatbaf, Ali and Moradi, Mehrdad and Almasizadeh, Jaafar and Sangchoolie, Behrooz and Van Acker, Bert and Denil, Joachim},
  booktitle={2022 18th European Dependable Computing Conference (EDCC)},
  pages={57--64},
  year={2022},
  organization={IEEE}
}

@inproceedings{garg2024coupling,
  title={On the Coupling between Vulnerabilities and LLM-generated Mutants: A Study on Vul4J dataset},
  author={Garg, Aayush and Degiovanni, Renzo and Papadakis, Mike and Le Traon, Yves},
  booktitle={2024 IEEE Conference on Software Testing, Verification and Validation (ICST)},
  pages={305--316},
  year={2024},
  organization={IEEE}
}

@article{khanfir2023efficient,
  title={Efficient mutation testing via pre-trained language models},
  author={Khanfir, Ahmed and Degiovanni, Renzo and Papadakis, Mike and Traon, Yves Le},
  journal={arXiv preprint arXiv:2301.03543},
  year={2023}
}

@article{cotroneo2024neural,
  title={Neural Fault Injection: Generating Software Faults from Natural Language},
  author={Cotroneo, Domenico and Liguori, Pietro},
  journal={arXiv preprint arXiv:2404.07491},
  year={2024}
}

@misc{repo,
  author = {Ahmad Hatahet},
  title = {LLMs-Powered Real-Time Fault Injection},
  url = {https://github.com/ahmadhatahet/hil-fault-test-gen},
  urldate = {2025-04-28}
}

@misc{ollama,
  author = {{Ollama}},
  title = {Ollama},
  url = {{https://github.com/ollama/ollama}},
  urldate = {2025-04-28}
}

@misc{llama3modelcard,
    title={Llama 3 Model Card},
    author={AI@Meta},
    year={2024},
    url = {https://github.com/meta-llama/llama3/blob/main/MODEL_CARD.md}
}

@article{qwen2,
      title={Qwen2 Technical Report}, 
      author={An Yang and Baosong Yang and Binyuan Hui and Bo Zheng and Bowen Yu and Chang Zhou and Chengpeng Li and Chengyuan Li and Dayiheng Liu and Fei Huang and Guanting Dong and Haoran Wei and Huan Lin and Jialong Tang and Jialin Wang and Jian Yang and Jianhong Tu and Jianwei Zhang and Jianxin Ma and Jin Xu and Jingren Zhou and Jinze Bai and Jinzheng He and Junyang Lin and Kai Dang and Keming Lu and Keqin Chen and Kexin Yang and Mei Li and Mingfeng Xue and Na Ni and Pei Zhang and Peng Wang and Ru Peng and Rui Men and Ruize Gao and Runji Lin and Shijie Wang and Shuai Bai and Sinan Tan and Tianhang Zhu and Tianhao Li and Tianyu Liu and Wenbin Ge and Xiaodong Deng and Xiaohuan Zhou and Xingzhang Ren and Xinyu Zhang and Xipin Wei and Xuancheng Ren and Yang Fan and Yang Yao and Yichang Zhang and Yu Wan and Yunfei Chu and Yuqiong Liu and Zeyu Cui and Zhenru Zhang and Zhihao Fan},
      journal={arXiv preprint arXiv:2407.10671},
      year={2024}
}

@article{phi4,
  title={Phi-4 technical report},
  author={Abdin, Marah and Aneja, Jyoti and Behl, Harkirat and Bubeck, S{\'e}bastien and Eldan, Ronen and Gunasekar, Suriya and Harrison, Michael and Hewett, Russell J and Javaheripi, Mojan and Kauffmann, Piero and others},
  journal={arXiv preprint arXiv:2412.08905},
  year={2024}
}

@misc{gpt4o,
  author = {{Azure AI Team}},
  title = {OpenAI’s fastest model, GPT-4o mini is now available on Azure AI},
  url = {{https://azure.microsoft.com/en-us/blog/introducing-gpt-4o-openais-new-flagship-multimodal-model-now-in-preview-on-azure/}},
  urldate = {2025-04-28}
}

@misc{gpt4o_mini,
  author = {{Azure AI Team}},
  title = {OpenAI’s fastest model, GPT-4o mini is now available on Azure AI},
  url = {{https://azure.microsoft.com}},
  urldate = {2025-04-28}
}

@article{fewshots,
  title={Language models are few-shot learners},
  author={Brown, Tom and Mann, Benjamin and Ryder, Nick and Subbiah, Melanie and Kaplan, Jared D and Dhariwal, Prafulla and Neelakantan, Arvind and Shyam, Pranav and Sastry, Girish and Askell, Amanda and others},
  journal={Advances in neural information processing systems},
  volume={33},
  pages={1877--1901},
  year={2020}
}

@article{amyan2024automating,
  title={Automating Fault Test Cases Generation and Execution for Automotive Safety Validation via NLP and HIL Simulation},
  author={Amyan, Ayman and Abboush, Mohammad and Knieke, Christoph and Rausch, Andreas},
  journal={Sensors},
  volume={24},
  number={10},
  pages={3145},
  year={2024},
  publisher={MDPI}
}

@article{abboush2024representative,
  title={Representative Real-Time Dataset Generation Based on Automated Fault Injection and HIL Simulation for ML-Assisted Validation of Automotive Software Systems},
  author={Abboush, Mohammad and Knieke, Christoph and Rausch, Andreas},
  journal={Electronics},
  volume={13},
  number={2},
  pages={437},
  year={2024},
  publisher={MDPI}
}

@article{cot,
  author       = {Jason Wei and
                  Xuezhi Wang and
                  Dale Schuurmans and
                  Maarten Bosma and
                  Ed H. Chi and
                  Quoc Le and
                  Denny Zhou},
  title        = {Chain of Thought Prompting Elicits Reasoning in Large Language Models},
  journal      = {CoRR},
  volume       = {abs/2201.11903},
  year         = {2022},
  url          = {https://arxiv.org/abs/2201.11903},
  eprinttype    = {arXiv},
  eprint       = {2201.11903},
  bibsource    = {dblp computer science bibliography, https://dblp.org}
}

@article{feedback_loop,
  title={Feedback loops with language models drive in-context reward hacking},
  author={Pan, Alexander and Jones, Erik and Jagadeesan, Meena and Steinhardt, Jacob},
  journal={arXiv preprint arXiv:2402.06627},
  year={2024}
}

@inproceedings{gpqa,
  title={Gpqa: A graduate-level google-proof q\&a benchmark},
  author={Rein, David and Hou, Betty Li and Stickland, Asa Cooper and Petty, Jackson and Pang, Richard Yuanzhe and Dirani, Julien and Michael, Julian and Bowman, Samuel R},
  booktitle={First Conference on Language Modeling},
  year={2024}
}

\end{document}